\documentclass[aps,prb,twocolumn,floats,epsfig]{revtex4}
\usepackage{amssymb}
\usepackage{amsbsy}
\usepackage{amsmath}
\usepackage{epsfig}
\usepackage{color}

\newcommand{\ing}{\includegraphics}
\newcommand{\bib}{\bibitem}
\newcommand{\beq}{\begin{equation}}
\newcommand{\eeq}{\end{equation}}
\newcommand{\bea}{\begin{eqnarray}}
\newcommand{\eea}{\end{eqnarray}}
\newcommand{\al}{\alpha}
\newcommand{\de}{\delta}
\newcommand{\De}{\Delta}

\newcommand{\ka}{\kappa}

\newcommand{\ta}{\theta}

\newcommand{\non}{\nonumber}

\begin{document}

\title{Transport across junctions of pseudospin-one fermions}

\author{Sourav Nandy$^{(1)}$, K. Sengupta$^{(1)}$, and Diptiman Sen$^{(2)}$}

\affiliation{$^{(1)}$School of Physical Sciences, Indian Association for the
Cultivation of Science, 2A and 2B Raja S. C. Mullick Road, Jadavpur 700032,
India \\
$^{(2)}$Centre for High Energy Physics, Indian Institute of Science,
Bengaluru 560012, India}

\date{\today}

\begin{abstract}

We study transport across ballistic junctions of materials which
host pseudospin-one fermions as emergent low-energy quasiparticles.
The effective low-energy Hamiltonians of such fermions are described
by integer spin Weyl models. We show that current
conservation in such integer spin-$s$ Weyl systems requires
continuity across a boundary of only $2s$ (out of $2s+1$) components
of the wave function. Using the current conservation conditions, we
study the transport between normal metal-barrier-normal metal (NBN)
and normal metal-barrier-superconductor (NBS) junctions of such
systems in the presence of an applied voltage $eV$.
We show that for a specific value of the barrier potential $U_0$,
such NBN junctions act as perfect collimators; any quasiparticle
which is incident on the barrier with a non-zero angle of incidence
is reflected back with unit probability for any barrier width $d$.
We discover an interesting symmetry of this system,
namely, the conductance is invariant under $U_0 \to 2(\mu_L \pm
eV)-U_0$, where $\mu_L$ is the chemical potential and the +(-) sign
corresponds to particle (hole) mediated transport. For NBS junctions
with a proximity-induced $s$-wave pairing potential, which also
display such a collimation, we chart out the properties of the
subgap tunneling conductance $G$ as a function of the barrier
strength and applied voltage. We point out the effect of the
collimation on the subgap tunneling conductance of these NBS
junctions and discuss experiments which can test our theory.

\end{abstract}

\maketitle

\section{Introduction}
\label{intro}

Symmetry protected touching of fermionic bands at isolated points in
the Brillouin zone leads to a rich class of phenomena in several
condensed matter systems~\cite{rev1}. In cases where such touching
occurs between the conduction and valence bands at the Fermi energy,
the effective low-energy fermions display emergent pseudospin
degrees of freedom representing band quantum numbers. When $2m+1$
such bands touch at the Fermi point, the low-energy effective theory
of such fermions is given by a spin-$m$ Weyl theory~\cite{rev2}.
For $m=1/2$, where two bands touch each other at the
Fermi surface, such systems represent Weyl semimetals. These
semimetals host several unconventional properties which distinguish
them from ordinary metals~\cite{weylpaper1,weylpaper2,weylpaper3}.

It is well-known that a touching of more than two bands at any given
point in the Brillouin zone is accidental. However, the presence of
additional symmetries may protect such a band touching under suitable
conditions~\cite{hasan1}. Examples of these are seen in several
symmorphic crystals which host mirror plane and discrete rotational
symmetries~\cite{hasan1}. It has been theoretically demonstrated,
via first principle calculations~\cite{bandcal1}, that three bands
may cross at the Fermi points in several symmorphic crystal systems
such as ${\rm MoP}$, ${\rm TiS}$, ${\rm RhSi}$, ${\rm TaN}$, and
${\rm ZrSe}$~\cite{crystalrefs}. These lead to the so called
triple-point or pseudospin-one fermion systems; the low-energy
effective Hamiltonian of such systems are described by an effective
pseudospin-one Weyl theory. We note that unlike spin-half Dirac or
Weyl fermion systems, integer pseudospin fermions have no analogs in
high-energy physics where their presence is naturally prohibited by the
spin-statistics theorem. Such band touchings can only occur in pairs
and can happen at specific points in the Brillouin zone whose
positions are dictated by the symmetries of the system. Keeping
these properties in mind a toy model having two such pseudospin-one
Weyl nodes has been put forward~\cite{stern1}.

Such pseudospin-one fermions host several unconventional features
that have no analogs in standard metals. First, the band touching
points or nodes act as a source or sink of Abelian Berry
curvature~\cite{bitan1}. For systems where the effective low-energy
dispersion of the fermions around the node goes as $\hbar v_F
\sqrt{k_{\perp}^{2n}\alpha_n^2 + k_z^2}$ (where $n$ is an integer,
$v_F$ is the Fermi velocity, $\hbar$ is Planck's constant, $\alpha_n$
is a constant, and $k_{\perp}=\sqrt{k_x^2+k_y^2}$), these nodes host
a topological charge of $2n$. Second, they host Fermi arcs on their
surfaces which have qualitatively distinct features from their
spin-half Weyl counterparts~\cite{arcref1}. Third, they are expected
to display large anomalous Hall conductivity and a quadratic
dependence of the magnetothermal conductivity on the external
magnetic field $B$ for small $B$. Moreover, in contrast to their
counterparts in half-integer Weyl and Dirac semimetals, such
fermions host a flat band at zero energy which makes them ideal
candidates for studying strong correlation physics.

The transport properties of fermions are well-known to provide
direct signatures of their topological nature. The simplest example
of this is the behavior of two-dimensional Dirac quasiparticles in
graphene in the presence of a barrier. In a ballistic normal
metal-barrier-normal metal (NBN) junction, which constitutes a
region with a barrier potential $U_0$ between two normal regions,
the tunneling conductance $G$ oscillates with $U_0$~\cite{novo1}.
This is in sharp contrast to the behavior of Schr\"odinger electrons
in such junctions where $G$ is a monotonic function of $U_0$.
Moreover such junctions allow for perfect transmission when either
an electron is incident on them normally or for specific values of
$U_0$ at any angle of incidence. The former phenomenon is known as
Klein tunneling and is a consequence of the inability of the barrier
to flip the electron spin (or pseudospin in the case of graphene) on
scattering. The latter phenomenon, known as transmission resonance,
occurs when the dimensionless barrier strength $\chi=U_0 d/(\hbar
v_F) = n \pi$, where $d$ is the barrier width and $n$ is an
integer. Both these features are distinct signatures of the Dirac
nature of the low-energy quasiparticles and are not seen in
conventional metals. Similar behavior can also be seen in normal
metal-barrier-superconductor (NBS) and
superconductor-barrier-superconductor (SBS) junctions of such
materials~\cite{ks1,ks2,been1,titov1,linder1}. These phenomena also
occur in NBN and NBS junctions of three-dimensional Weyl and multi-Weyl
semimetals~\cite{zhang1}. Moreover, it was recently pointed out that
the tunneling conductance $G$, in NBN and NBS junctions between a
Weyl and a multi-Weyl semimetal with different winding numbers,
becomes independent of the barrier strength for sufficiently thin
barriers~\cite{sinha1}. It was shown that such a barrier independence
is a consequence of the change of the topological winding number across
the junction. However, the transport features of NBN and NBS junctions
involving pseudospin-one fermions have not yet been studied.

In this work, we study transport across ballistic NBN and NBS
junctions whose basic quasiparticles are pseudospin-one fermions.
The central results of our study are as follows. First, we show that
for any integer pseudospin $s$ fermion system, current conservation
in NBN (NBS) junctions require continuity of only $2s ~(4s)$ out of
the $2s+1 ~(4s+2)$ components of the fermion wave function. This
feature is unique to integer pseudospin fermions; for Weyl or Dirac
fermions with half-integer spin current conservation necessarily
implies continuity of the entire wave function. Second, we
demonstrate the presence of perfect collimation in such NBN
junctions. We find that when the barrier potential $U_0$ is tuned
such that $U_0 = \mu_L +(-) eV$ (where $\mu_L$ is the chemical
potential, $eV$ is the applied voltage across the junction, and the
$+(-)$ sign corresponds to particle (hole) mediated transport), such
junctions become completely opaque to all incident fermions
approaching the barrier region at non-zero angles of incidence. In
contrast, fermions which approach the junction at normal incidence
are transmitted with unit probability. We demonstrate that this
collimation occurs for any width $d$ of the barrier region which
makes them distinct from analogous behavior in junctions hosting
spin-half Weyl or Dirac fermions. We tie the presence of such
collimation to the lack of continuity of some of the components of
the wave function across the junction and note that it makes such
junctions ideal test beds for studying Klein tunneling. Third, we
unravel a symmetry property of $G$ in such junctions; we note that
$G$ remains invariant under $U_0 \to 2(\mu_L +(-) eV)-U_0$ for
particle (hole) mediated transport, and this invariance does not
depend on chemical potential or topological winding number
differences across the junction. Fourth, we study the tunneling
conductance of such NBN junctions and demonstrate that they display
an oscillatory behavior as a function of the barrier potential
$\chi$ if the topological winding number does not change across the
junction; in contrast, $G$ becomes independent of $\chi$ if the
winding number changes. Finally, we study the behavior of the subgap
tunneling conductance across a NBS junction of such a material. To
this end, we envisage a simple model where $s$-wave
superconductivity with a pairing amplitude $\Delta$ is induced by a
proximate superconductor, and we study the behavior of the subgap
tunneling conductance as a function of the barrier potential and
applied voltage across the junction. Our analysis
reveals an approximate symmetry of the subgap tunneling conductance
under the transformation $\delta \to -\delta$, where
$\delta=U_0-\mu_L$, for any applied voltage $eV \le \Delta$. We also
chart out the signature of collimation in the subgap tunneling
conductance and discuss experiments which can test our theory.

The plan of the rest of the paper is as follows. In Sec.~\ref{nbn}
we discuss transport through NBN junctions and discuss collimation
in these junctions. This is followed by a study of subgap tunneling
conductance in NBS junctions in Sec.~\ref{nbs}. Finally, we
summarize our results, discuss experiments which can test our
theory, and conclude in Sec.~\ref{diss}.

\section{NBN junction}
\label{nbn}

\begin{figure}
\centering {\ing[width=\linewidth]{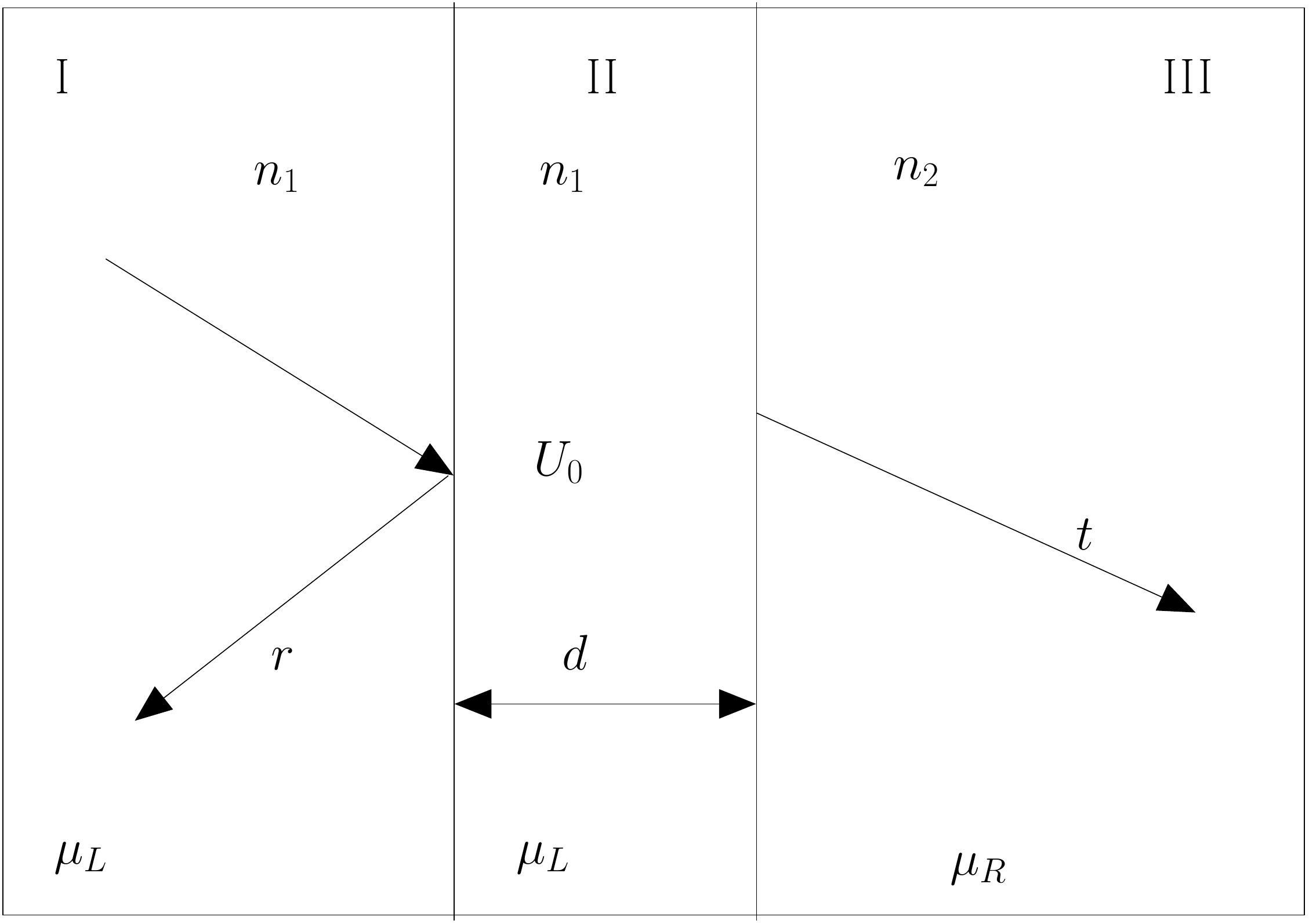}}
\caption{Schematic picture of transmission through a ballistic NBN
junction hosting pseudospin-one low-energy quasiparticles. The longitudinal
coordinate $z$ increases from left to right, and the barrier region (region
$II$) with a potential $U_0$ has a width $d$ along $z$. $\mu_L$ and $\mu_R$
denote the chemical potentials in regions $I$ and $III$ respectively.}
\label{fig1} \end{figure}

In this section, we will analyze transport through an NBN junction
of pseudospin-one fermions. The setup is schematically shown in
Fig.~\ref{fig1}. The normal regions $I$ and $III$ host pseudospin-one
fermions whose Hamiltonian is $H= \sum_{\vec k} \psi^{\dagger}_{\vec k}
H_n(\vec k) \psi_{\vec k}$, where $\psi_{\vec k}$ is a three-component
fermion field and $H_n(\vec k)$ is given by
\begin{eqnarray} \mathcal H_n(\vec k) ~=~ \sum_{a=x,y,z} ~d_{a n}(\vec k) ~
S_a, \label{Ham} \end{eqnarray}
$d_{1,2,3 n}(\vec k)$ are functions of momenta as specified below, and
$S_x$, $S_y$ and $S_z$ are the generators of $S=1$ algebra given by
\begin{eqnarray} S_x &=& \frac{1}{\sqrt{2}} \begin{pmatrix}
0 & 1 & 0 \\
1 & 0 & 1 \\
0 & 1 & 0 \end{pmatrix},~~
S_y=\frac{1}{\sqrt{2}} \begin{pmatrix}
0 & -i & 0 \\
i & 0 & -i \\
0 & i & 0
\end{pmatrix}, \non \\
S_z &=& \begin{pmatrix}
1 & 0 & 0 \\
0 & 0 & 0 \\
0 & 0 & -1 \end{pmatrix}. \end{eqnarray}
The functions $d_{1,2,3 n}(\vec k)$ depend on the winding number
$n$ of the Weyl nodes. In pseudospin-one fermions systems,
combinations of different symmetries usually restrict $n \le 3$. For $n=1$,
these functions are given by~\cite{bitan1}
\begin{eqnarray} d_{11}(\vec k)=v_F k_x, \quad d_{21}(\vec k)=v_F k_y, \quad
d_{31}(\vec k)=v_F k_z, \end{eqnarray}
where $v_F $ is the Fermi velocity. For $n=2$, we have
\begin{eqnarray} d_{12}(\vec k) &=& \al_2(k_x^2-k_y^2),\quad d_{22}(\vec
k)=2\al_2k_yk_x, \non \\
d_{32}(\vec k)&=& v_F k_z. \end{eqnarray}
Here $\al_2/\hbar^2$ has the dimension of inverse mass. For $n=3$, we have
\begin{eqnarray} d_{13}(\vec k) &=& \al_3(k_x^3-3k_y^2k_x),\quad d_{23}(\vec k)
=\al_3 (k_y^3-3k_yk_x^2), \non \\
d_{33}(\vec k) &=& v_F k_z. \end{eqnarray}
A straightforward diagonalization of $H_n(\vec k)$ leads to
\begin{eqnarray} E^n_{\pm}(\vec k) &=& \pm \sqrt{[d_{1n}(\vec k)]^2+
[d_{2n}(\vec k)]^2+[d_{3n}(\vec k)]^2},\non \\
E^n_0 (\vec k) &=& 0, \end{eqnarray}
where $+(-)$ indicates the conduction (valence) band, and $E_0^n$
represents the flat band at zero energy.

To study transport, we analyze the passage of an electron through the junction
with an energy $\mu_L + eV >0$, where $eV$ is the applied voltage and $\mu_L$
is the chemical potential. To this end, we first find
the eigenstate for the positive energy band in region $I$ where the
Weyl nodes have a topological winding number $n_1$. This is most
easily done via the following coordinate transformations:
\begin{eqnarray} k_x &=& \Big(\frac{\mu_L + eV}{\al_{n_1}}\sin\ta_{1\vec k}
\Big)^{1/n_1}\cos \phi_{\vec k}, \\
k_y &=& \Big(\frac{\mu_L + eV}{\al_{n_1}}\sin\ta_{1\vec k}\Big)^{1/n_1}
\sin\phi_{\vec k}, \non \\
k_{1z} &=& (\mu_L + eV) \cos\ta_{1 \vec k}/v_F, \non \end{eqnarray}
where $\phi_{\vec k}=\tan^{-1}(k_y/k_x)$, $\ta_{1 \vec k}=\arccos[v_F k_{1z}/
(\mu_L + eV)]$, and $k_{\perp}=\sqrt{k_x^2+k_y^2}$.
In terms of $\ta_{1\vec k}$ and $\phi_{\vec k}$, the eigenstate
for a right [R] (left [L]) moving fermion having $k_z
>(<) 0$ is given by:
\begin{eqnarray} |\psi^{n_1}\rangle_R &=& e^{i( -n_1 S_z \phi_{\vec k} +k_x x
+ k_y y + k_z z)} \begin{pmatrix}
\cos^2\frac{\ta_{1\vec k}}{2} \\
\frac{\sin\ta_{1\vec k}}{\sqrt{2}} \\
\sin^2\frac{\ta_{1\vec k}}{2} \end{pmatrix}, \label{wavfnr} \\
|\psi^{n_1}\rangle_L &=& e^{i( -n_1 S_z \phi_{\vec k} +k_x x + k_y y
- k_z z)} \begin{pmatrix}
\sin^2\frac{\ta_{1\vec k}}{2} \\
\frac{\sin\ta_{1\vec k}}{\sqrt{2}} \\
\cos^2\frac{\ta_{1\vec k}}{2} \end{pmatrix}. \label{wavfnl} \end{eqnarray}
Note that the $\phi_{\vec k}$-dependence of the wave function can be
envisaged as a rotation around the $\hat z$ axis in spin space by an angle
$n_1 \phi_{\vec k}$. In terms of these eigenfunctions, the wave function in
region $I$ can be written as
\begin{eqnarray} |\psi\rangle_I &=& |\psi^{n_1}\rangle_R + r |\psi^{n_1}
\rangle_L, \label{wav1} \end{eqnarray}
where $r$ denotes the reflection amplitude.

In region $II$, the wave function $|\psi\rangle_{II}$ consists of right and
left-moving fermions,
\begin{eqnarray} |\psi\rangle_{II} &=& p |\psi^{'n_1}\rangle_R + q |\psi^{'n_1}
\rangle_L, \label{wav2} \end{eqnarray} where $p$ and $q$ denote the
amplitudes of right- and left-moving fermions respectively, and the
wave functions $|\psi^{'n_1}\rangle_{R,L}$ are given by
Eqs.~\eqref{wavfnr} and \eqref{wavfnl} with $\ta_{1 \vec k}$ and
$k_{1z}$ replaced by
\begin{eqnarray} \ta_{1 \vec k} \to \ta_{2 \vec k}&=& \arccos[v_F k_{2z}/
(\mu_L + eV -U_0)] \label{ta} \\
k_{1z} \to k_{2z} &=& \sqrt{(\mu_L + eV -U_0)^2/(\hbar v_F)^2 - \al_{n_1}^2
k_\perp^{2n_1}}. \non
\end{eqnarray}

Finally in region $III$, the transmitted fermion has a wave function
\begin{eqnarray} |\psi\rangle_{III} &=& t |\psi^{n_2}\rangle_R, \label{wav3}
\end{eqnarray}
where $t$ is the transmission amplitude, and the wave function
$|\psi^{n_2}\rangle_R$ for a right-moving fermion in region $III$ is
given by Eq.~\eqref{wavfnr} with $n_1$ replaced by $n_2$,  $\ta_{1
\vec k} \to \ta_{3\vec k} = \arccos[v_F k_{3z}/(\mu_R + eV)]$, and
$k_{1z} \to k_{3z} = \sqrt{(\mu_R + eV)^2/(\hbar v_F)^2 -
\al_{n_2}^2 k_\perp^{2n_2}}$.

Next, to determine $r$, $t$, $p$ and $q$, we follow the standard
procedure of imposing current conservation at $z=0$ and $z=d$. We
note that there are four complex coefficients to be determined.
However, a continuity of the entire wave function, which usually
follows from current conservation in junctions hosting Dirac or Weyl
fermions with linear dispersion, would lead to six equations. The
solution to this conundrum comes from noticing that the current
$J_{\al} \sim \psi^{\dagger} S_{\al} \psi$ for pseudospin-one
fermions always involves only two of the three components of the
wave function. This is most easily seen for the current along $z$;
if the fermion wave function is given by $\psi \sim
(c_1,c_2,c_3)^T$, we have $J_z \sim |c_1|^2 -|c_3|^2$ which does not
involve $c_2$. Thus current conservation in these junctions do not
require continuity of all the components of the wave function across
the junction. In the rest of this work, we will focus on the current
along $z$ and impose current conservation by demanding continuity of
only the first and the third components of the wave function which
appear in the expression of $J_z$. We do not impose any restriction
on the second component of the wave function which does not appear
in $J_z$. We note that this property of pseudospin-one fermions
follows from the non-invertibility of the spin matrices $S_{\al}$
and is therefore not shared by their half-integer-spin counterparts.
Although it is not directly relevant to our study here, we would
like to note that such a discontinuity of the wave function is a
general property of integer pseudospin Dirac/Weyl fermions; for any
integer pseudospin $s$, current conservation would require
continuity of only $2s$ of the $2s+1$ components of the wave function.

The equations obtained by imposing continuity of the first and the third
component of the wave functions (which, as discussed above, is sufficient
for ensuring current continuity along $z$) at $z=0$ and $z=d$ can be read
off from Eqs.~\eqref{wav1}, \eqref{wav2}, and \eqref{wav3}. We find that
\begin{widetext}
\begin{eqnarray} \label{BC1}
\cos^2\frac{\ta_{1 \vec k}}{2} ~+~ r\sin^2\frac{\ta_{1 \vec k}}{2} &=&
p \cos^2 \frac{\ta_{2 \vec k}}{2} ~+~ q \sin^2\frac{\ta_{2 \vec k}}{2}, \\
\sin^2\frac{\ta_{1 \vec k}}{2} ~+~ r\cos^2\frac{\ta_{1 \vec k}}{2} &=&
p \sin^2\frac{\ta_{2 \vec k}}{2}~+~ q \cos^2\frac{\ta_{2 \vec k}}{2}, \non \\
p \cos^2\frac{\ta_{2 \vec k}}{2}e^{ik_{2z} d} ~+~ q \sin^2
\frac{\ta_{2 \vec k}}{2}e^{-ik_{2z} d}&=& t\cos^2\frac{\ta_{3 \vec k}}{2}
e^{i(k_{3z}d + \nu)}, \non \\
p \sin^2\frac{\ta_{2 \vec k}}{2}e^{ik_{2z} d} ~+~ q \cos^2
\frac{\ta_{2 \vec k}}{2}e^{-ik_{2z} d} &=& t\sin^2\frac{\ta_{3 \vec k}}{2}
e^{i(k_{3z} d -\nu)}, \non \end{eqnarray}
\end{widetext}
where $\nu= (n_1 - n_2)\phi_{\vec k}$. The transmission probability $T=
1-|r|^2$ can be found by solving for $r={\mathcal N}/{\mathcal D}$ from
Eqs.~\eqref{BC2} where
\begin{widetext}
\begin{eqnarray} \mathcal N &=& \sin^2\left(\frac{\ta_2}{2}\right) \cos^2\left(
\frac{\ta_3}{2}\right) e^{2 i \nu \phi} \left(\left(\cos
\left(\ta_1\right) +1\right) \left(\cos
\left(\ta_2\right)+1\right)-2 \left(\cos \left(\ta_1
\right)+\cos \left(\ta_2\right)\right) e^{2 i d k_{2 z}}\right) \non \\
&&- ~\sin^2\left(\frac{\ta_3}{2}\right) \left(4
\sin^4\left(\frac{\ta_2}{2} \right)
\cos^2\left(\frac{\ta_1}{2}\right)+\sin^2\left(\frac{\ta_1}{2}
\right)
\sin^2\left(\ta_2\right) \left(-1+e^{2 i d k_{2 z}}\right)\right) \non \\
&&+ ~\cos^4\left(\frac{\ta_2}{2}\right) \left(4
\sin^2\left(\frac{\ta_3}{2} \right)
\cos^2\left(\frac{\ta_1}{2}\right) e^{2 i d k_{2 z}}-4 \sin^2\left(
\frac{\ta_1}{2}\right) \cos^2\left(\frac{\ta_3}{2}\right) e^{2 i \nu
\phi}
\right), \non \\
\mathcal D &=& 4 \cos^2\left(\frac{\ta_1}{2}\right)
\cos^4\left(\frac{\ta _2}{2}\right)
\cos^2\left(\frac{\ta_3}{2}\right) e^{2 i \nu \phi} + e^{i \left(2 d
k_{2 z}+\nu \phi \right)}\cos \left(\ta_1\right)-\cos \left(\ta_2
\right)\left(\left(\cos \left(\ta_2\right)-\cos
\left(\ta_3\right)\right)
\cos (\nu \phi )\right. \non \\
&& \left.+ ~i \left(\cos \left(\ta_2\right) \cos
\left(\ta_3\right)-1\right) \sin (\nu \phi
)\right)-\sin^2\left(\frac{\ta_1}{2}\right) \sin^2\left(
\ta_2\right) \cos^2\left(\frac{\ta_3}{2}\right) e^{2 i \nu \phi} \non \\
&& + ~4 \sin^2\left(\frac{\ta_1}{2}\right)
\sin^4\left(\frac{\ta_2}{2}\right)
\sin^2\left(\frac{\ta_3}{2}\right)-\sin^2\left(\ta_2\right)
\sin^2\left( \frac{\ta_3}{2}\right)
\cos^2\left(\frac{\ta_1}{2}\right). \label{BC2} \end{eqnarray}
\end{widetext}
Note that we have now omitted the momentum index $\vec k$ for
$\ta_{1,2,3}$ for clarity. The conductance $G$ can be computed
from the transmission probability in a straightforward manner using
the standard Landauer-Buttiker prescription by summing over all the
transmission channels~\cite{lbref}. We then find
\begin{eqnarray}
\label{int} G &=& G_0\int_0^{\pi/2}d\ta_1\int_0^{2\pi}d\phi_k ~J_0 ~T, \\
G_0 &=& \frac{n_0 e^2 k_F^2 L^2}{h N_1},\quad
N_1=\int_0^{\pi/2}d\ta_1 \int_0^{2\pi}d\phi_k ~J_0=\pi n_1, \non
\end{eqnarray} where $J_0$ denotes the Jacobian of the
transformation from $(k_x,k_y)$ to $(\ta_1, \phi_k)$ and is given by
$J_0=\sin(\ta_1)^{(2-n_1)/n_1}\cos(\ta_1)$, and $n_0$ is the total
number of Weyl nodes. Here $G_0$ measures the number of available
channels at all Weyl nodes, $k_F$ is the Fermi wave vector, and $L^2
\equiv L_x L_y$ denote the transverse dimensions of the sample. We
have assumed here the absence of internode scattering upon
reflection from the barrier. Such an assumption can be justified in the
case where the Weyl nodes occur at different transverse momenta since
scattering from the barrier must respect transverse momentum conservation.

\begin{figure}
\centering {\ing[width=\linewidth]{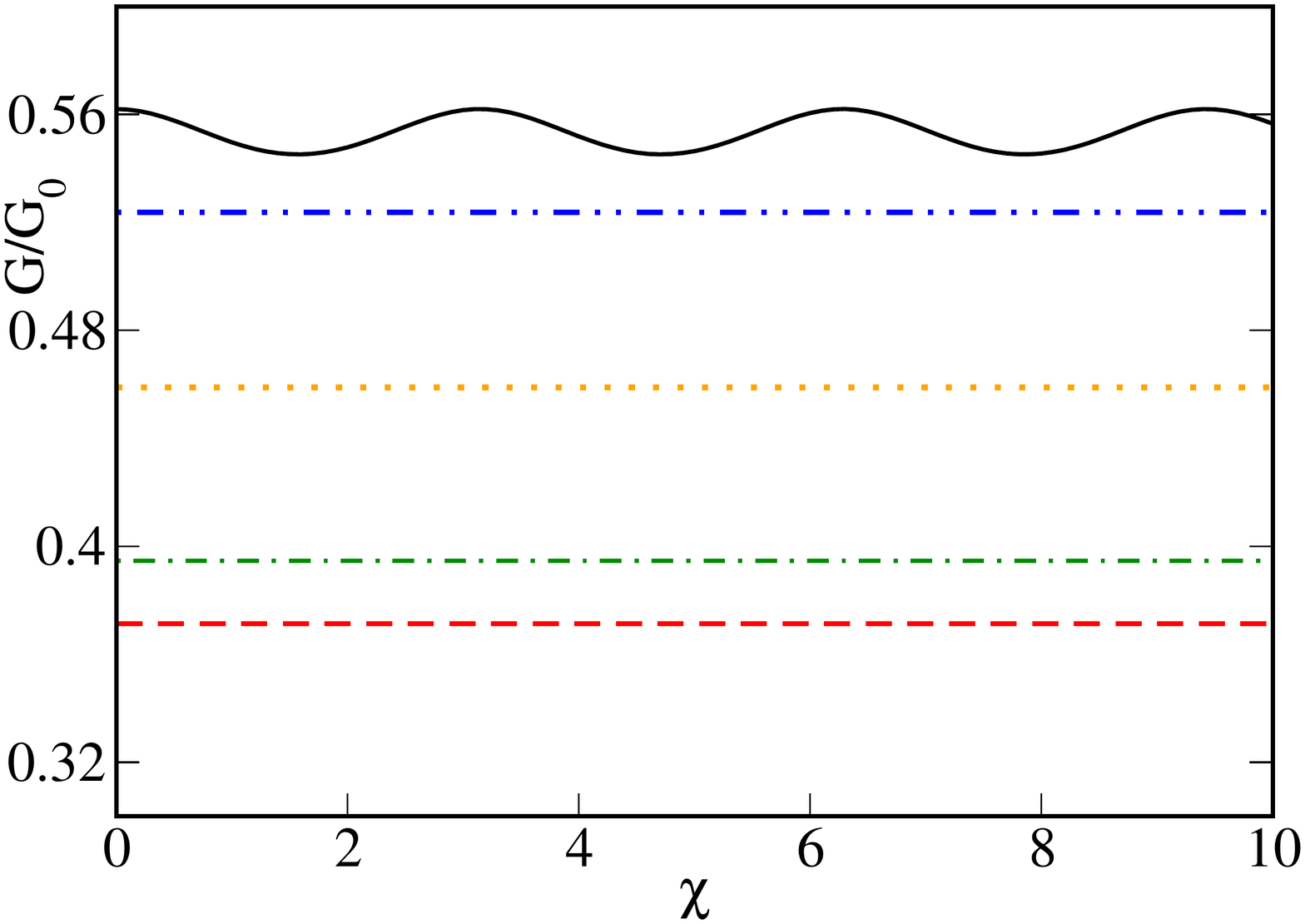}} \caption{Plot of
$G/G_0$ for $\de n= 0$ (black solid line), $\pm 1$ (red dashed line
for $-1$ and yellow dotted line for $1$), and $\pm 2$ (green
dash-dotted line for $-2$ and blue dash-double dotted line for $2$)
indicating oscillatory behavior (for $\de n=0$) or constant behavior
(for $\de n= \pm 1$ and $\pm 2$) as a function of $\chi$.
For all the curves, we have chosen $n_2=1$ for $\delta n =0,1,2$
and $n_1=1$ for $\delta n=-1,-2$. Here we have set $\mu_R=0.5$,
$\mu_L= \hbar v_F k_F=1$, $eV=0.2$. All energies are scaled in units
of $\mu_L$.} \label{fig2} \end{figure}

\begin{figure}
\centering {\ing[width=\linewidth]{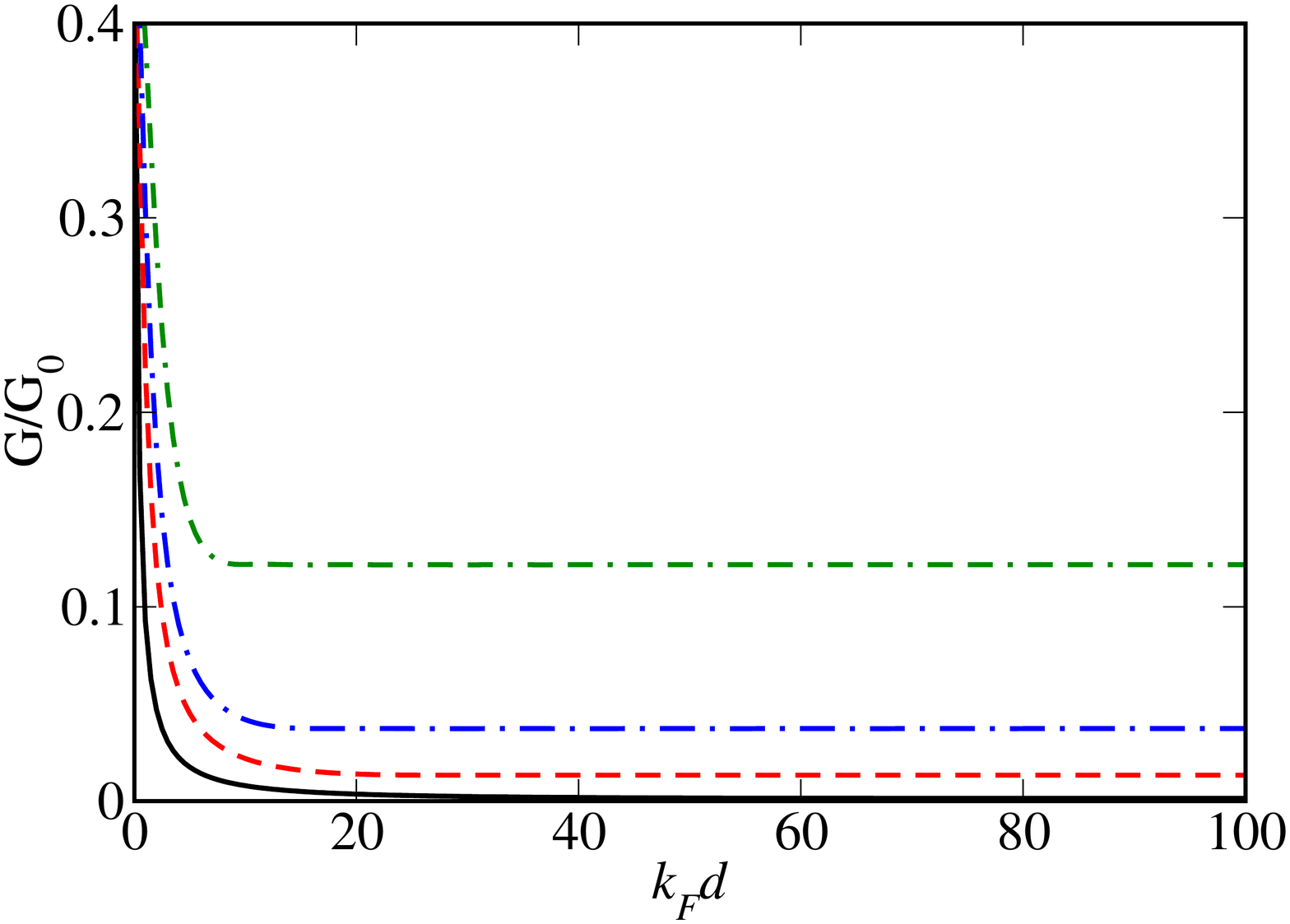}} \caption{Plot of
$G/G_0$ as a function of $d$ for several representative values of
$U_0$. Here $\mu_R=0.5$ and all other parameters are the same as in
Fig.~\ref{fig2}. The black solid line corresponds to
$\mu_L + eV -U_0=0.05$, the red dashed to $\mu_L + eV -U_0=0.15$, the blue
dash-dotted to $\mu_L +eV -U_0=0.35$, and the green double dash-dotted
to $\mu_L +eV -U_0=0.45$. Here $d$ is measured in units of $\hbar
v_F/\mu_L = k_F^{-1}$ and all energies are measured in units of
$\mu_L$.} \label{fig3} \end{figure}

To make further analytical progress, we consider the thin barrier
limit in which $U_0 \to \infty$ and $d \to 0$ keeping $\chi= U_0
d/\hbar v_F$ fixed. In this limit $\ta_{2 \vec k}, k_{3z} d \to
0$ and $k_{2 z} d \to \chi$. Using this one obtains from Eqs.~\eqref{BC1}
\begin{eqnarray} r &=& \frac{\sin \left(\frac{\ta_3}{2}\right) \cos
\left(\frac{\ta_1}{2}\right)-\sin \left(\frac{\ta_1}{2}\right) \cos \left(
\frac{\ta_3}{2}\right) e^{2 i \nu'}}{\cos \left(\frac{\ta_1}{2}\right) \cos
\left(\frac{\ta _3}{2}\right) e^{2 i \nu'}- \sin \left(\frac{\ta_1}{2}\right)
\sin \left(\frac{\ta_3}{2}\right)}, \end{eqnarray}
where $\nu'= (n_1-n_2)\phi_k - \chi$. Thus we find that in this
limit, for $n_1 \ne n_2$, the barrier potential $\chi$ appears as a
constant shift to the azimuthal angle $\phi_k$. Consequently, $G$
which involves a sum over all such angles becomes independent of
$\chi$. In contrast, for $n_1= n_2$, $G$ is an oscillatory
function of the barrier potential. These properties of $G$ in these
junctions are qualitatively similar to those found in ballistic
junctions of spin-half Weyl semimetals~\cite{sinha1}. This behavior
is numerically confirmed in Fig.~\ref{fig2} where $G/G_0$ is
plotted as a function of $\chi$ for $\de n \equiv n_1-n_2= 0, \pm 1,
\pm 2$. We find that $G/G_0$ oscillates with $\chi$ for $\de n=0$;
in contrast it is independent of $\chi$ for $\de n \ne 0$. This
independence persists for a wide range of $d$ and $U_0$ even when we move
away from the restrictive thin barrier limit as shown in Fig.~\ref{fig3}.

\begin{figure}
\centering {\ing[width=\linewidth]{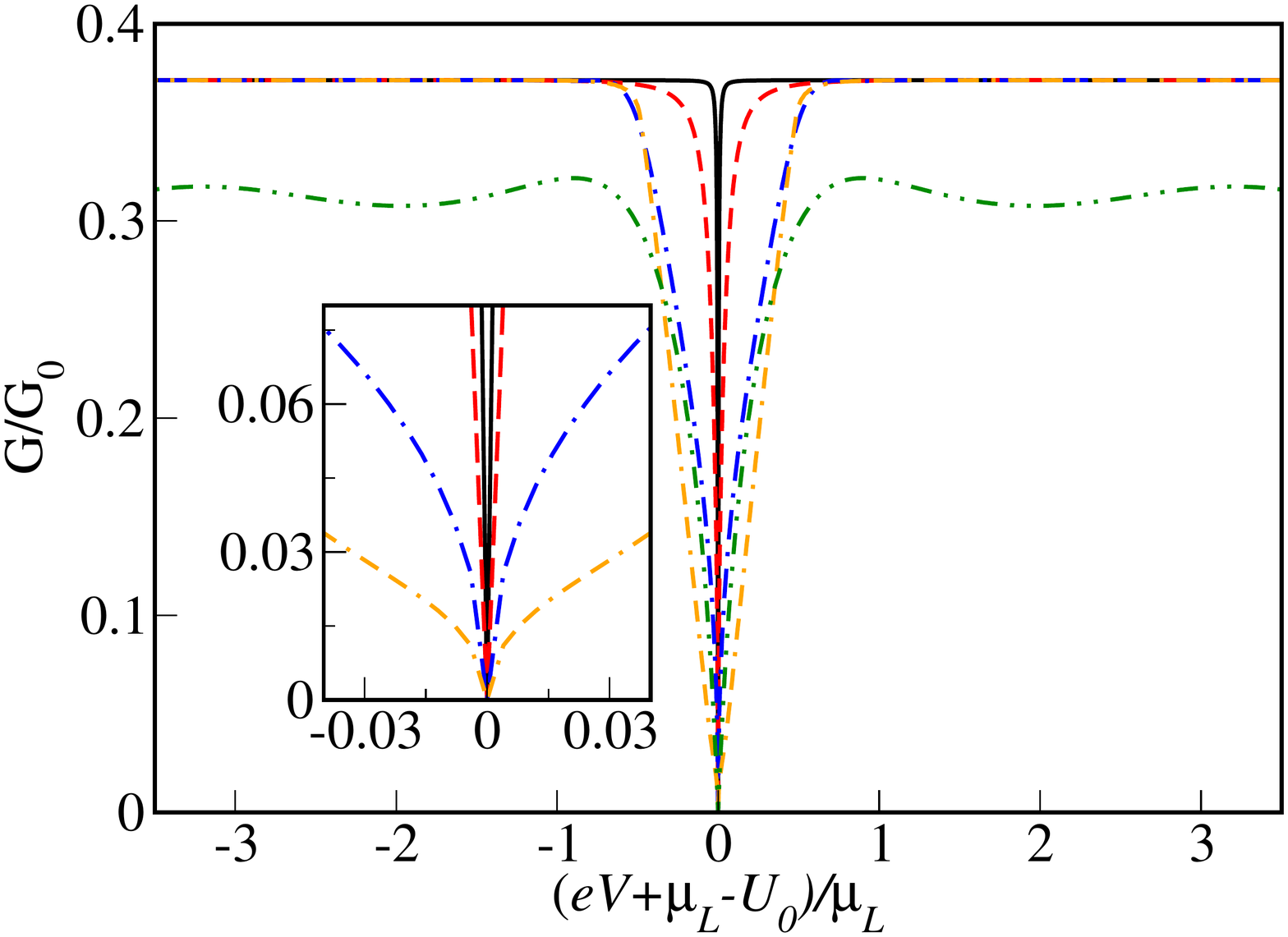}} \caption{Plot of
$G/G_0$ as a function of $\mu_L +eV-U_0$ for several values of $\de
n$ and $d$, with a fixed applied voltage $eV=0.2$ and chemical
potentials $\mu_L=1$ and $\mu_R=0.5$. The green dotted line
corresponds to $dk_F=1$ and $\de n = 0$. The orange double
dash-dotted, the blue dash-dotted, the red dash-dotted and the black
solid lines correspond to $\de n=-1$ and $d k_F=50$, $d k_F=5$, $d
k_F=1$ and $d k_F=0.5$ respectively. The convention for 
choosing $n_1$ and $n_2$ for a given $\delta n$ is the same as in 
Fig.~\ref{fig2}. All energies are in units of $\mu_L$. The inset presents a 
closer view of $G/G_0$ around $\mu_L +eV-U_0=0$.} \label{fig4} \end{figure}

Next, we study the $U_0$ dependence of the barrier potential close
to $U_0 = \mu_L + eV$ where we deviate significantly from the thin
barrier limit for any $d$. The result is shown in Fig.~\ref{fig4}.
Remarkably, we find that $G$ vanishes for $U_0 =\mu_L + eV$ for all
values of $d$ and $\de n$. The approach of $G/G_0$ to its zero value
depends on $d$ for a given $eV$; a thicker barrier region with a
larger value of $d$ leads to a more gradual decay of $G/G_0$ as can
be seen in Fig.~\ref{fig4}. We have also checked that this property
is independent of $eV$. We note that this property is distinct from
the analogous behavior of $G/G_0$ for two-component Dirac and Weyl
fermions; for these materials $G/G_0$ does not approach zero for
sufficiently thin barriers even if $U_0 = \mu_L +eV$.

To understand this phenomenon better, we first note that for $\mu_L +eV -U_0
=0$, the wave function in region $II$ must satisfy the equation
\begin{eqnarray} \sum_{a=x,y,z} S_a d_{a n} (\vec k) \psi_0(\vec k) = 0
\label{hamzero} \end{eqnarray} for any $k_{\perp} \ne 0$ and $k_z$.
This requires imaginary solutions for $k_{2z} = \pm i \ka_n $ where
$ \ka_n = k_{\perp}^n/\al_n$. This leads to evanescent modes in
region $II$. The wave functions of these modes can be found by
solving Eq.~\eqref{hamzero}. The wave function in region $II$ is
thus given by
\begin{eqnarray} |\psi^{(0)}\rangle_{II} &=& \frac{1}{\sqrt{3}} \Bigg[ p
e^{-\ka_n z + i(k_x + k_y y - n S_z \phi_k)} \begin{pmatrix}
1\\
i\\
-1 \end{pmatrix} \non \\
&& + q e^{-\ka_n z + i(k_x + k_y y - n S_z \phi_k)} \begin{pmatrix}
-1\\
i \\
1 \end{pmatrix} \Bigg]. \label{wav0} \end{eqnarray}
The wave functions in region $I$ and $III$ do not involve $U_0$ and are given
by Eqs.~\eqref{wav1} and \eqref{wav3} respectively. Using these wave functions
and matching the first and the third components of the wave function as before,
we obtain
\begin{eqnarray} &&\cos^2\frac{\ta_{1 \vec k}}{2}+r\sin^2\frac{\ta_{1 \vec
k}}{2}=(p-q)\sqrt{3}, \non \\
&&\sin^2\frac{\ta_{1 \vec k}}{2}+r\cos^2\frac{\ta_{1 \vec
k}}{2}= (q - p)/\sqrt{3}, \non \\
&& p e^{-\ka_n d}- q e^{\ka_n d}= \sqrt{3} t\cos^2\frac{\ta_{3 \vec k}}{2}
e^{i(k_{3z}d + \nu)}, \non \\
&& q e^{\ka_n d}-p e^{-\ka_n d}=\sqrt{3} t\sin^2\frac{\ta_{3 \vec
k}}{2} e^{i(k_{3z} d -\nu)}. \label{bczero} \end{eqnarray} The only
possible solution to Eqs.~\eqref{bczero} is $r=-1$, $t=0$ and $p/q=
\exp[2 \ka_n d]$ which indicates perfect reflection of electrons for
all $k_{\perp} \ne 0$. We note that this phenomenon
is independent of $d$ and $\mu_R$; moreover it can occur at any value
of $eV$ and $\mu_L$ provided the condition $U_0 = \mu_L +eV$ is satisfied.

In contrast, for $k_{\perp}=0$, i.e., when the particle approaches the
barrier at normal incidence, the wave function in region $II$ is given by
\begin{eqnarray} |\psi^{'(0)}\rangle_{II} &=& p e^{-\ka_n z + i(k_x + k_y y
- n S_z \phi_k)} \begin{pmatrix}
1\\
0\\
0 \end{pmatrix} \non \\
&& + ~q e^{-\ka_n z + i(k_x + k_y y - n S_z \phi_k)} \begin{pmatrix}
0\\
0 \\
1 \end{pmatrix}. \label{wav00} \end{eqnarray}
Then a straightforward calculation yields $r=0$ and $t=1$ for any
$d$. This is of course a manifestation of the well-known Klein
tunneling. Thus we find that the barrier for $U_0= \mu_L +eV$ reflects
all electrons with unit probability except the ones which are incident
on it normally; the latter are transmitted with unit probability.
This leads to perfect collimation in such NBN junctions. This also
explains the reason for $G/G_0 \to 0$ in this limit; $G/G_0$ is suppressed
by a factor of $1/L^2$ factor since only one of the channels conduct.

When both $\De E \equiv \mu_L +eV - U_0$ and the angle of incidence
(or, equivalently, $k_\perp$) are close to zero, there is a
cross-over from perfect reflection to perfect transmission (Klein
tunneling) as the angle of incidence approaches zero.
For $ \hbar v_F \alpha_{n_1}k_{\perp}^{n_1}, \De E \ll
(\mu_L +eV), (\mu_R +eV)$, where $\theta_{1,3} \to 0$ but $\theta_2$
remains finite, we find analytically that the transmission
probability is given by \beq T ~=~ \left[1 ~+~ \left( \frac{\hbar
v_F d \alpha_{n_1}^2 k_\perp^{2n_1}}{2 \De E} \right)^2\right]^{-1}.
\label{cross} \eeq The above expression holds for any $n_2$ and
implies that the width $\De E$ of the cross-over region is
proportional to $d$. This explains why the width of the region of
small $G/G_0$ in Fig.~\ref{fig4} decreases as $d$ becomes smaller.
Using Eq.~\eqref{cross} and the Jacobian $J_0$ in Eq.~\eqref{int}
to integrate over $\theta_1$, we find that $G/G_0 \sim
(\De E/d)^{1/n_1}$ for $\De E \ll (\mu_L +eV)$. In principle,
this scaling form gives a way of experimentally measuring the value
of $n_1$, although it may be very hard to study the region of small
$\De E$ since $G$ would be small in this regime.

Before ending the discussion of the collimation effect, we would
like to note that this is rather unique to integer pseudospin
fermion systems since it can only occur in a system where current
conservation does not enforce continuity of the entire wave
function. This can be seen by noting that the fermion wave function
in region $III$, $|\psi\rangle_{III}$, vanishes for any $k_{\perp}
\ne 0$ since $t=0$ for $\mu_L + eV= U_0$. In contrast, in region
$II$ the wave function $|\psi^{(0)}\rangle_{II}$ is finite and is
given by Eq.~\eqref{wav0} with $p/q= \exp[2 \ka d]$. Thus this
solution necessarily requires a wave function discontinuity at
$z=d$. Also, it is easy to see using similar analysis that for hole
mediated transport an analogous collimation would occur at $U_0 = \mu_L -eV$.

We now turn to an opposite effect called super-Klein
tunneling~\cite{xu}. For $U_0= 2(\mu_L +eV)$, we find from
Eqs.~\eqref{BC1} that the transmission probability for a given
incident momentum $\vec k$ is given by \beq T ~=~ \frac{4 \cos
\ta_{1 \vec k} \cos \ta_{3 \vec k}}{(\cos \ta_{1 \vec k} + \cos
\ta_{3 \vec k})^2 ~+~ 4 \sin^2 \nu \sin^2 \ta_{1 \vec k} \sin^2
\ta_{3 \vec k}}. \eeq We now see that if $n_1 = n_2$ and $\mu_L =
\mu_R$, $T=1$ for all values of $\ta_{1 \vec k}$; this is called
super-Klein tunneling. However, we see that this phenomenon does not
occur if either $n_1 \ne n_2$ or $\mu_L \ne \mu_R$.

\begin{figure}
\centering {\ing[width=\linewidth]{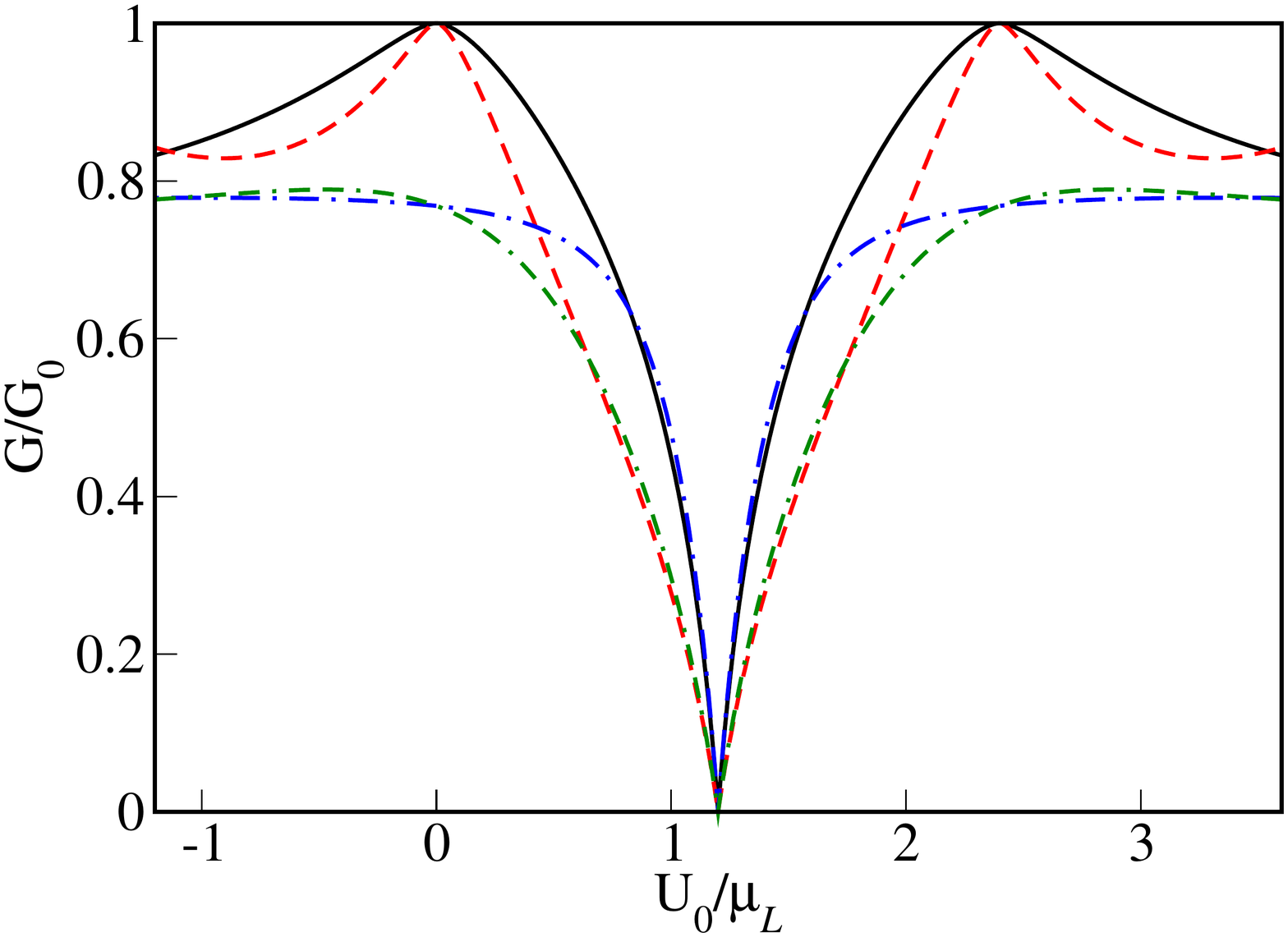}} \caption{Plot of
$G/G_0$ as a function of $U_0$ for several values of $\de n$ and
$d$, with a fixed applied voltage $eV=0.2$ and chemical potentials
$\mu_L = \mu_R = 1$. The black solid, red dashed, blue dash-dotted
and green double dash-dotted lines correspond to $(d k_F,\de n)$
equal to $(0.5,0), ~(1,0), ~ (0.5,-1)$ and $(1,-1)$
respectively. The convention for choosing $n_1$ and
$n_2$ for a given $\delta n$ is the same as in Fig.~\ref{fig2}. All
energies are in units of $\mu_L$.} \label{fig5} \end{figure}

Finally, we would like to point out a remarkable symmetry of the
conductance as a function of $U_0$ for any value of $d$ and $\de n$,
namely, that $G/G_0$ has the same value at two values of the barrier
potential $U_0$ which are related to each other by reflection about
the value $\mu_L +eV$. This is clearly visible in Fig.~\ref{fig5}
where $\mu_L +eV= 1.2$; reflection about $\mu_L +eV$ then
corresponds to the values $U_0$ and $2.4 - U_0$. To show this symmetry,
suppose that Eqs.~\eqref{BC1} describe the various amplitudes for a
value $U_0$. Then we find that at $2(\mu_L +eV) -U_0$, the
corresponding equations (with amplitudes denoted by primes) are given by
\begin{widetext}
\begin{eqnarray} \label{BC3}
\cos^2\frac{\ta_{1 \vec k}}{2}~+~ r' \sin^2\frac{\ta_{1 \vec k}}{2} &=&
q' \cos^2 \frac{\ta_{2 \vec k}}{2} ~+~p' \sin^2\frac{\ta_{2 \vec k}}{2}, \\
\sin^2\frac{\ta_{1 \vec k}}{2} ~+~r' \cos^2\frac{\ta_{1 \vec k}}{2} &=&
q' \sin^2\frac{\ta_{2 \vec k}}{2} ~+~ p' \cos^2\frac{\ta_{2 \vec k}}{2}, \non \\
q' \cos^2\frac{\ta_{2 \vec k}}{2}e^{ik_{2z} d} ~+~ p' \sin^2
\frac{\ta_{2 \vec k}}{2}e^{-ik_{2z} d} &=& t' \cos^2\frac{\ta_{3 \vec k}}{2}
e^{i(k_{3z}d + \nu')}, \non \\
q' \sin^2\frac{\ta_{2 \vec k}}{2}e^{ik_{2z} d} ~+~ p' \cos^2
\frac{\ta_{2 \vec k}}{2}e^{-ik_{2z} d} &=& t' \sin^2\frac{\ta_{3 \vec k}}{2}
e^{i(k_{3z} d -\nu')}. \non \end{eqnarray}
\end{widetext}
We now observe that complex conjugating Eqs.~\eqref{BC1} precisely
give Eqs.~\eqref{BC3} if we take $p'=q^*$, $q'=p^*$, $r'=r^*$, $t' =
t^* e^{-i2k_{3z}d}$, and $\nu' = - \nu$. These relations mean that
the transmission probability is the same (i.e., $1 - |r|^2 = 1 -
|r'|^2$) for the values $U_0$ and angle $\phi_{\vec k}$ and the
values $2(\mu_L +eV) -U_0$ and angle $- \phi_{\vec k}$. Since the
conductance is calculated by integrating over all angles $\phi_{\vec
k}$ from 0 to $2\pi$ (equivalently, from $-\pi$ to $\pi$), we see
that the conductance will be the same for $U_0$ and $2(\mu_L +eV)
-U_0$. Clearly, this argument holds for any value of $d$ and $\de
n$. Also, it is easy to see that identical arguments
would hold for hole mediated transport; however, in that case, $G$
would be invariant under $U_0 \to 2(\mu_L -eV)-U_0$.

\section{NBS junction}
\label{nbs}

\begin{figure}
\centering {\ing[width=\linewidth]{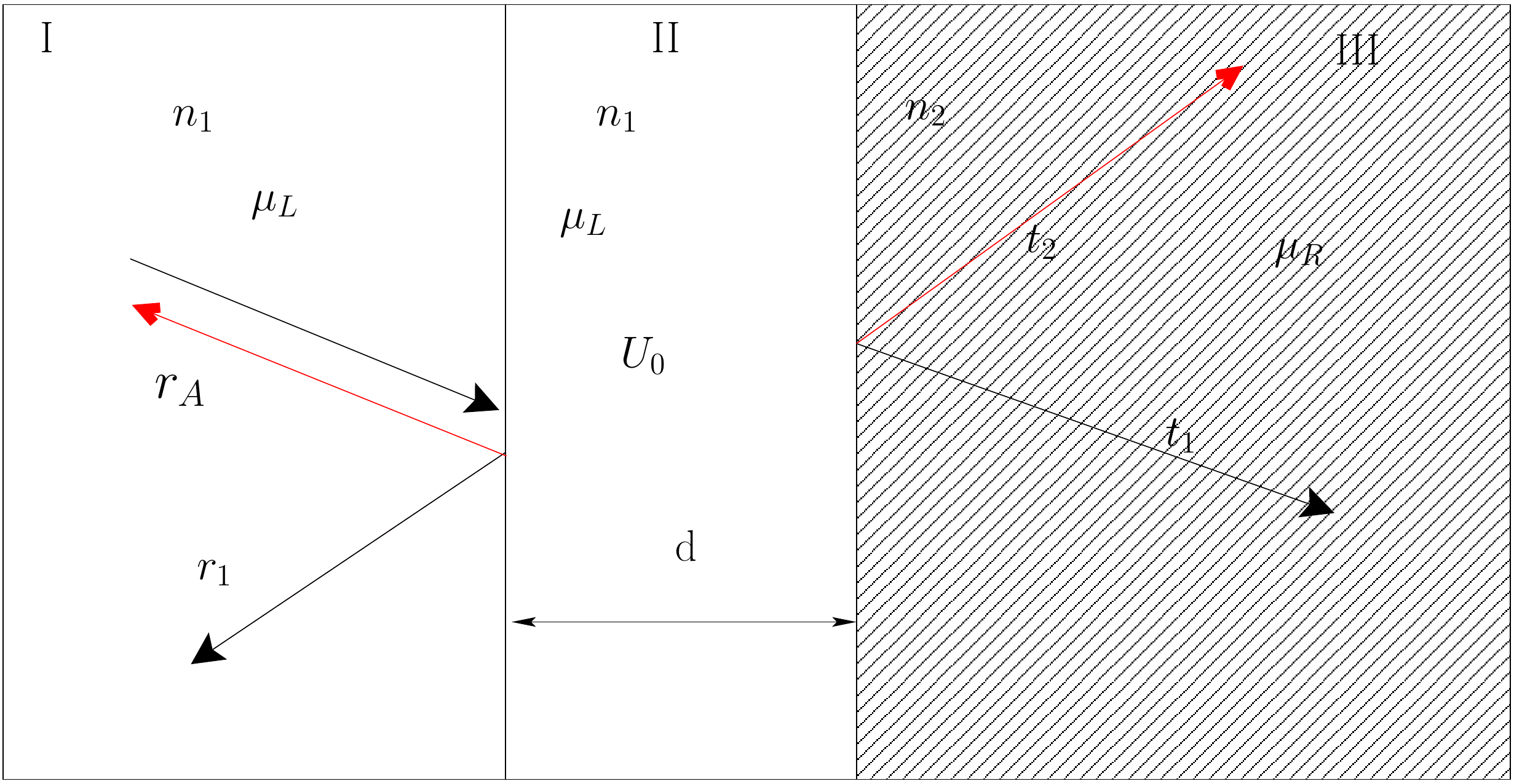}} \caption{Schematic
picture of a ballistic NBS junction with a proximate superconductor
(not shown in the figure) atop region $III$ (shaded region). The
longitudinal coordinate is $z$ and the barrier region (region $II$)
with a potential $U_0$ has a width $d$ along $z$. $\mu_L$ and
$\mu_R$ denote the chemical potentials in regions $I$ and $III$
respectively. The figure shows the amplitudes of both normal ($r$)
and Andreev reflections ($r_A$) in region I. $t_1$ and $t_2$ denotes
the amplitudes of electron- and hole-like quasiparticles in the
superconducting region.} \label{fig6} \end{figure}

In this section, we will study transport through a NBS junction
which hosts pseudospin-one fermions. A schematic picture of the
proposed setup is shown in Fig.~\ref{fig6}. In what follows, we
assume that superconductivity is induced in region $III$ by a
proximate $s$-wave superconductor leading to the induction of a $s$-wave
pair potential between two Weyl nodes $A$ and $B$. We note that
several possibilities of unconventional (i.e., non-$s$-wave)
superconductivity have recently been proposed as possible phases of
pseudospin-one fermion systems~\cite{lin1}; however, in this
work we will assume $s$-wave symmetry. The Hamiltonian of the
system in the presence of such $s$-wave pairing in region $III$ is given by
\begin{eqnarray} H_s &=& \sum_{\vec k} ~\psi_{\vec k}^{\dagger} ~[\tau_z
(H_{n_2}(\vec k) -\mu_R) + \tau_x \De]~ \psi_{\vec k} \non \\
&=& \sum_{\vec k} ~\psi_{\vec k}^{\dagger} H_s(\vec k) \psi_{\vec k},
\label{supham} \end{eqnarray}
where $H_{n_2}(\vec k)$ is the normal state Hamiltonian defined in
Eq.~\eqref{Ham}, $\tau_{z,x}$ are spin-half Pauli matrices in
particle-hole space, and $\psi_{\vec k}$ is a six-component spinor
field whose top three components represent pseudospin-one electron
wave functions at Weyl node $A$ while the bottom three components represent
hole wave functions at node $B$. The eigenfunctions corresponding to
right-moving electron- and hole-like quasiparticles obtained by solving
$H_s(\vec k) \psi_{\vec 3k}= (\mu_R +eV) \psi_{\vec k}$ are given by
\begin{widetext}
\begin{eqnarray} \psi_{3e}^s &=& e^{i(k_{3z} z +k_x x + k_y y)} \exp[-i S_z
\phi_{\vec k}] \left( e^{i\beta}\cos^2\frac{\ta_3}{2}, e^{i\beta}
\frac{\sin\ta_3}{\sqrt{2}}, e^{i\beta}\sin^2\frac{\ta_3}{2}, \cos^2
\frac{\ta_3}{2}, \frac{\sin\ta_3}{\sqrt{2}}, \sin^2\frac{\ta_3}{2}\right),
\non \\
\psi_{3h}^s &=& e^{-i(k'_{3z} z +k_x x + k_y y)} \exp[-i S_z \phi_{\vec k}]
\left(\sin^2\frac{\ta_3^{\prime}}{2}, \frac{\sin\ta_3^{\prime}}{\sqrt{2}},
\cos^2\frac{\ta_3^{\prime}}{2}, e^{i\beta}\sin^2\frac{\ta_3^{\prime}}{2},
e^{i\beta}\frac{\sin\ta_3^{\prime}}{\sqrt{2}}, e^{i\beta}\cos^2
\frac{\ta_3^{\prime}}{2}\right), \label{supwav} \end{eqnarray}
\end{widetext}
where
$\ta_3(\ta_3^{\prime})=\arctan [\al_{n_2}k_{\perp}^{n_2}/k_{3e}^{z+}
(k_{3h}^{z+})]$ and $k_{3z}(k'_{3z})=\sqrt{(\mu_R
+(-)\Omega)^2-\al_{n_2}^2 k_{\perp}^{2n_2}}$. Here for $eV>\De$, we
have $\Omega=[(eV)^2- \De^2]^{1/2}$ and $\beta=-i{\rm arccosh}
(eV/\De)$ whereas, for $eV<\De$, $\Omega=i[\De^2-(eV)^2]^{1/2}$ and
$\beta=\arccos(eV/\De)$. In what follows, we will set the phase of
the superconductor pair potential to be zero and omit the $\vec k$
index for $\ta_3$, $\ta'_3$ and $\beta$ for clarity.

In region $I$, $\De=0$, and electron (hole) wave functions can be
obtained from solution of $\pm H_{n_1}(\vec k) \psi= (eV \pm \mu_L)
\psi$. We note that in region $I$, corresponding to a right-moving incident
electron on the barrier at $z=0$, there is a reflected left-moving
electron and an Andreev reflected left-moving hole. The
wave functions of these electrons and holes are given by
\begin{widetext}
\begin{eqnarray} \psi_{1R}^e &=& e^{ i ( k_{1z} z + k_x x + k_y y)} \exp[-i
n_1 S_z \phi_k] \left( \cos^2(\ta_1/2), \sin(\ta_1)/\sqrt{2}, \sin^2
(\ta_1)/2,0,0,0\right), \non \\
\psi_{1L}^e &=& e^{ i (- k_{1z} z + k_x x + k_y y)} \exp[-i n_1 S_z
\phi_k] \left( \sin^2(\ta_1/2), \sin(\ta_1)/\sqrt{2}, \cos^2(\ta_1)/2,0,0,0
\right), \non \\
\psi_{1L}^h &=& e^{ i ( -k'_{1z} z + k_x x + k_y y)} \exp[-i n_1 S_z
\phi_k] \left( 0,0,0,\cos^2(\ta'_1/2), -\sin(\ta'_1)/\sqrt{2},
\sin^2(\ta'_1)/2 \right), \label{ehwav1} \end{eqnarray}
\end{widetext}
where $\ta_1=\arcsin [\al_{n_1}k^{n_1}_{\perp}/|\mu_L +eV|]$, $\ta_1^{\prime}
=\arcsin [-\sin\ta_1 |\mu_L +eV|/|\mu_L - eV|]$, $k_{1z} =(\mu_L +eV)\cos\ta_1$,
and $k'_{1z}=(eV-\mu_L)\cos\ta_1^{\prime}$. The wave function in
region $I$ is thus given by
\begin{eqnarray} \psi_I^s &=& \psi_{1R}^e + r \psi_{1L}^e + r_A \psi_{1L}^h,
\label{wavs1} \end{eqnarray}
where $r(r_A)$ is the amplitude of normal (Andreev) reflection from the
barrier.

Similarly, in region $II$ the wave function consists of a linear
superposition of left- and right-moving electron and hole wave
functions. The wave functions for right- and left-moving electrons
and that of the left-moving hole are denoted by $\psi_{2R}^e$,
$\psi_{2L}^e$, and $\psi_{2L}^h$ respectively. Their expressions can
be read off from Eqs.~\eqref{ehwav1} with
\begin{eqnarray}
&& \ta_1 \to \ta_2= \arcsin [\al_{n_1}k^{n_1}_{\perp}/|\mu_L +eV -U_0|,
\nonumber\\ && \ta'_1 \to \ta'_2= \arcsin \left[-\sin\ta_2
\frac{|\mu_L +eV -U_0|}{|\mu_L - eV - U_0|}\right], \nonumber\\
&& k_{1z} \to k_{2z} =(\mu_L +eV -U_0)\cos\ta_2, \nonumber\\
&& k'_{1z} \to k'_{2z}=(eV-\mu_L +U_0)\cos\ta_2^{\prime}.
\label{reg2eqs} \end{eqnarray}
The wave function of the right-moving hole in region $II$ is given by
\begin{eqnarray} \psi_{2R}^h &=& e^{ i ( k'_{2z} z + k_x x + k_y y)} \exp[-i
n_1 S_z \phi_k] \label{hwav2} \\
&& \times \left( 0,0,0,\sin^2(\ta'_2/2), -\sin(\ta'_2)/\sqrt{2},
\cos^2(\ta'_2)/2 \right). \non \end{eqnarray}
The wave function in region $II$ can be written as a superposition of
these wave functions as
\begin{eqnarray} \psi_{II}^s &=& p_1 \psi_{2R}^e + q_1 \psi_{2L}^e + p_2
\psi_{2L}^h + q_2 \psi_{2R}^h. \label{wavs2} \end{eqnarray}

The wave function in region $III$ can be written as a linear superposition of
electron- and hole-like quasiparticle wave functions given in
Eq.~\eqref{supwav} and are given by
\begin{eqnarray} \psi_{III} &=& t_1 \psi_{3e}^s + t_2 \psi_{3h}^s.
\label{wavs3} \end{eqnarray}
Here $t_1$ and $t_2$ denotes amplitudes of electron- and
hole-like quasiparticles in $\psi_{III}$ respectively.

To compute the conductance $G$ of the NBS junction, we first need to
determine the coefficients $r$ and $r_A$. To this end, we demand
current conservation at $z=0$ and $z=d$. We find that similar to the
NBN junction, the current through NBS junctions of pseudospin-one
fermions do not involve all the components of the wave function;
consequently, the conservation does not necessitate continuity of
the entire wave function across the boundaries between region $I$
and $II$ and between $II$ and $III$. We note that for
NBS junctions hosting integer pseudospin $s$ fermions with $4s+2$
component wave functions, only $4s$ components of the wave function
would be continuous. Moreover, it is easy to see from
Eqs.~\eqref{supwav} and \eqref{ehwav1} that current conservation
along $z$ does not involve the second and the fifth components of
the wave functions in regions $I$, $II$ and $III$. Thus we enforce
the current conservation along $z$ by only demanding continuity of
the other four components of the wave function. The procedure is
similar to that charted out for NBN junctions and yields, at $z=0$,
\begin{eqnarray} \cos^2(\ta_1/2) + r \sin^2(\ta_1/2) &=& p_1 \cos^2(\ta_2/2)
+ q_1 \sin^2(\ta_2/2), \non \\
\sin^2(\ta_1/2) + r \cos^2(\ta_1/2) &=& p_1 \sin^2(\ta_2/2) + q_1 \cos^2
(\ta_2/2), \non \\
r_A \cos^2 (\ta'_1/2) &=& p_2 \cos^2 (\ta'_2/2) + q_2 \sin^2 (\ta'_2/2),
\non \\
r_A \sin^2 (\ta'_1/2) &=& p_2 \sin^2 (\ta'_2/2) + q_2 \cos^2 (\ta'_2/2).
\label{z0cond} \non \\
\end{eqnarray}
Similarly, at $z=d$ one obtains
\begin{eqnarray} && p_1 \cos^2(\ta_2/2)e^{i k_{2z} d} + q_1 \sin^2(\ta_2/2)
e^{-i k_{2z} d} \non \\
&&= [t_1 e^{i(\beta+ k_{3z}d)} \cos^2(\ta_3/2) + t_2 e^{-i k'_{3z}d} \sin^2
(\ta'_3/2)]e^{i \nu'}, \non \\
&& p_1 \sin^2(\ta_2/2)e^{i k_{2z} d} + q_1 \cos^2(\ta_2/2) e^{-i k_{2z} d}
\non \\
&&= [t_1 e^{i(\beta+ k_{3z}d)} \sin^2(\ta_3/2) + t_2 e^{-i k'_{3z}d}
\cos^2(\ta'_3/2)]e^{-i \nu'}, \non \\
&&p_2 \cos^2 (\ta'_2/2) e^{-i k'_{2z} d} + q_2 \sin^2 (\ta'_2/2) e^{i k'_{2z}
d} \label{zdcond} \\
&&= [t_1 e^{ik_{3z}d} \cos^2(\ta_3/2) + t_2 e^{-i (k'_{3z}d -\beta)}
\sin^2(\ta'_3/2)]e^{i \nu'}, \non \\
&&p_2 \sin^2 (\ta'_2/2) e^{-i k'_{2z} d} + q_2 \cos^2 (\ta'_2/2)
e^{i k'_{2z} d} \non \\
&&= [t_1 e^{ik_{3z}d} \sin^2(\ta_3/2) + t_2 e^{-i (k'_{3z}d -\beta)}
\cos^2(\ta'_3/2)]e^{-i \nu'}, \non
\end{eqnarray}
where $\nu'= (n_1-n_2)\phi_k$. From Eqs.~\eqref{z0cond} and \eqref{zdcond}
we solve numerically for $r$ and $r_A$. The conductance $G$ is then obtained
from the usual Landauer-Buttiker approach~\cite{btk1}
\begin{eqnarray} G_s(eV) &=& G_N \int d^2 k \, T_s, \quad T_s= (1-|r|^2
+|r_A|^2), \non \\
\label{supcondcal} \end{eqnarray}
where $G_N =n_0 e^2
(k_F L)^2 /h$ is the normal state conductance of region $I$, where
$k_F= \mu_L+eV/\hbar v_F$, and we have chosen $n_1=1$.
We note here that the range of the integration over the transverse
momentum $(k_x,k_y)$ in Eq.~\eqref{supcondcal} is determined by demanding that
$\theta_{1,2}$ and $\theta'_{1,2}$ in regions I and II have real
solutions \cite{ks1}.

\begin{figure}
\centering {\ing[width=\linewidth]{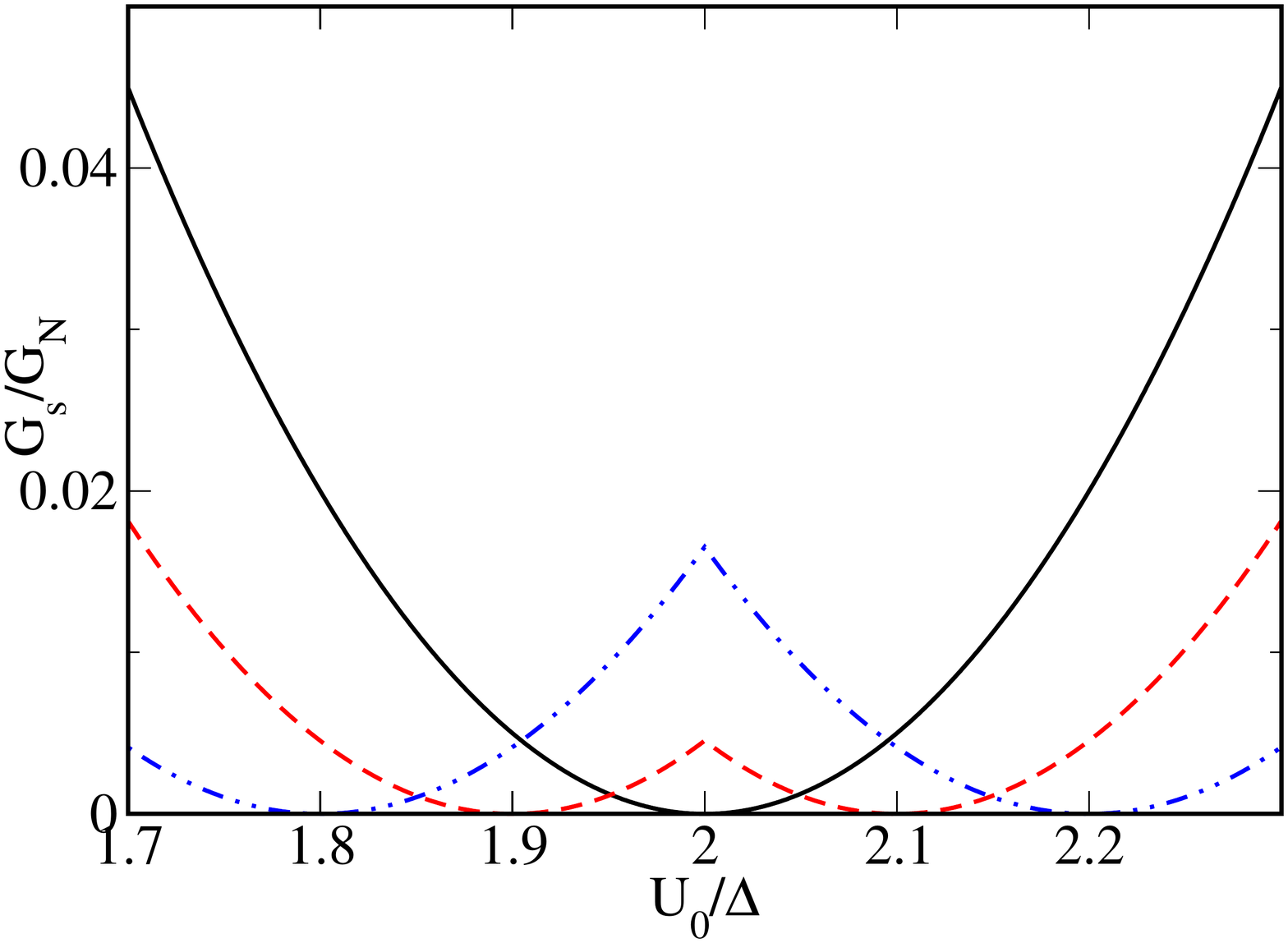}} \caption{Plot of
$G_s/G_N$ as a function of $U_0$ for $\mu_L=\mu_R=2$, $n_1=1$, $n_2=2$, and
$\De=1$. The black dash-dotted line corresponds to $eV=0$, the red dashed
line to $eV=0.1$, and the blue solid line to $eV=0.2$. The dips in $G_s$
occur when $U_0 = \mu_L + eV$ or $U_0 = \mu_L -eV$. All energies are scaled in
units of $\De$ and $dk_F=1$ for all plots.} \label{fig7} \end{figure}

We will first study the subgap conductance $G_s(eV)$ as a function
of $U_0$ near $U_0 = \mu_L + eV$. We note that in the NBN junction
when $U_0$ is tuned to this value, it led to collimation and a
consequent suppression of $G$. We find a similar suppression of
$G(eV)$ as shown in Fig.~\ref{fig7}. We find that $G_s(eV) \simeq 0$
for $U_0 = \mu_L + eV$ and its approach to zero as $ U_0 \to \mu_L +
eV$ is controlled by the thickness $d$ of the barrier region
(similar to the discussion around Eq.~\eqref{cross} for a NBN
junction). This behavior can be understood by noting that for
$k_{\perp} \ne 0$ and $U_0=\mu_L +eV$, the right- and the
left-moving electron wave functions in region $II$ are given by
Eq.~\eqref{wav0}. Thus the first two equations in
Eqs.~\eqref{z0cond} reduce to the first two equations in
Eqs.~\eqref{bczero}; they lead to a solution $r=-1$. Similarly the
first two equations in Eq.~\eqref{zdcond} can be shown to lead to
the solution $t_1=t_2=0$, similar to that found in last two
equations in Eqs.~\eqref{bczero}. This forces $p_2=q_2=r_A=0$
leading to a complete suppression of transmission for any non-zero
angle of incidence. This feature is reflected in the dips at
$U_0=\mu_L + eV$ (at $U_0/\Delta=2,\,2.1,\,2.2$) in Fig.~\ref{fig7}.
A similar argument can be given for $U_0= \mu_L -eV$ by considering
a hole approaching the barrier leading to a reflected hole with
amplitude $r$ and an Andreev reflected electron with amplitude
$r_A$. Once again a similar calculation to the one carried out above
shows that $r=-1$ and $r_A=0$ for $U_0=\mu_L -eV$. This leads to the
dips in $G_s$ for $U_0=\mu_L -eV$ at $U_0/\Delta = 2,\,1.9,\,1.8$.
These dips therefore constitute a concrete signature of collimation
in the subgap tunneling conductance of such NBS junctions.

\begin{figure}
\centering {\ing[width=\linewidth]{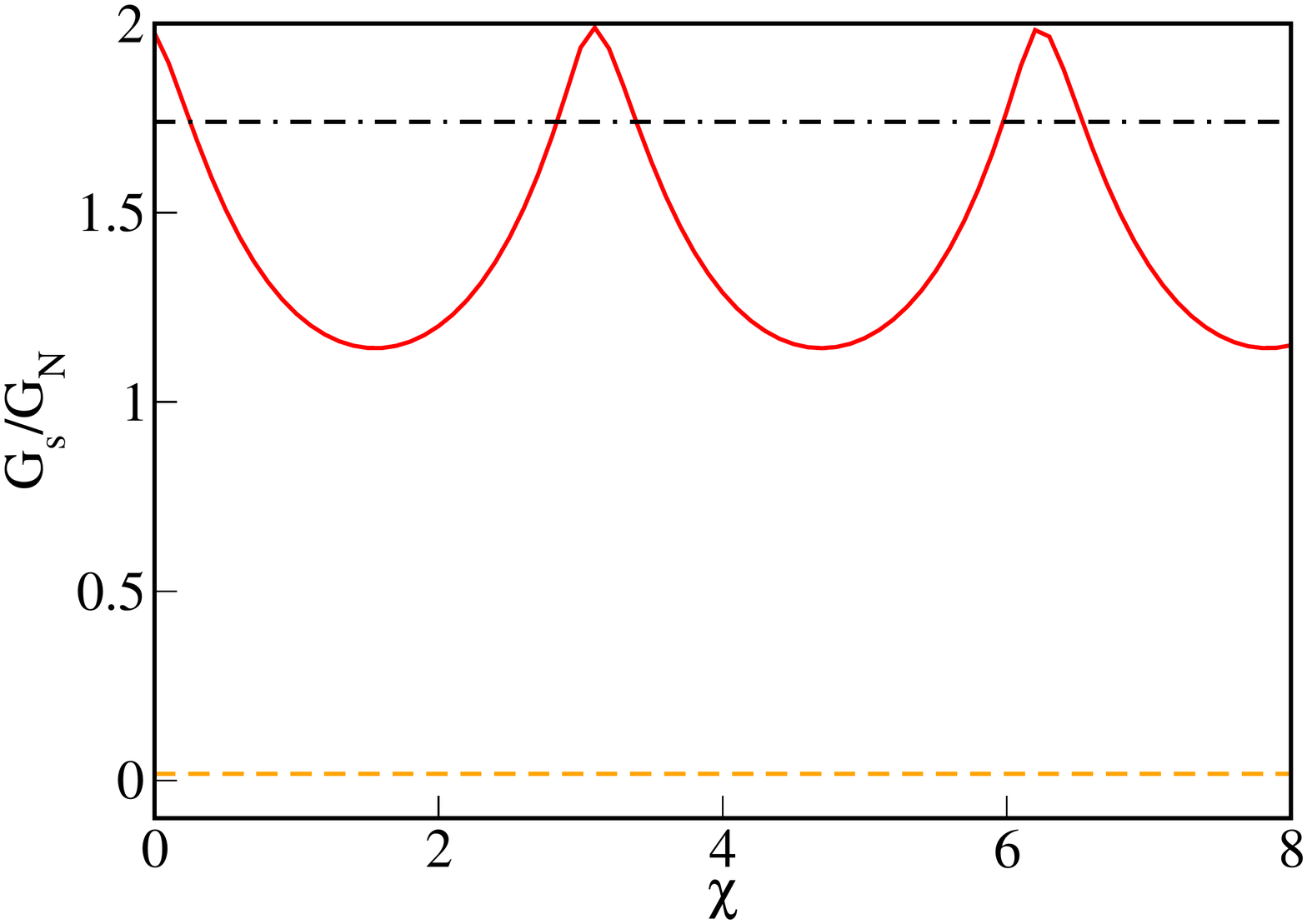}} \caption{Plot of the zero-bias 
tunneling conductance $G_s(0)/G_N$ as a function of $\chi$ for $eV=0, ~\mu_L=
\mu_R=100$ and $\De=1$. All energies are scaled in units of $\De$. The red 
solid line corresponds to $\de n=0$, the yellow dashed line to $\de n=-1$, 
and the black dash-dotted line to $\de n=1$. The convention 
for choosing $n_1$ and $n_2$ for a given $\delta n$ is the same as in 
Fig.~\ref{fig2}.} \label{fig8} \end{figure}

We note that Fig.~\ref{fig7} seems to indicate that
$G_s$ is reflection symmetric about $U_0 =\mu_L$. This is however only an
approximate symmetry which can be understood as follows. We first
note that in the parameter regime where $eV \ll \mu_L$ and $U_0$, $G_s$
receives contributions from only near-normal angles of incidence.
Indeed, in all the curves in Fig.~\ref{fig8}, the maximum angle of incidence
for which the channels conduct is given by $\theta^{\rm max} \le 0.15$,
and $\theta_1, \theta'_1 \le \theta^{\rm max}$. Thus we can replace
$\cos(\theta_1/2), \cos(\theta'_1/2) \to 1$ and $\sin(\theta_1/2),
\sin(\theta'_1/2) \to 0$ in Eqs.~\eqref{z0cond}. This leads to
\begin{eqnarray} 1 &\simeq & p_1 \cos^2(\ta_2/2) + q_1 \sin^2(\ta_2/2), \non \\
r &\simeq& p_1 \sin^2(\ta_2/2) + q_1 \cos^2 (\ta_2/2), \non \\
r_A &\simeq& p_2 \cos^2 (\ta'_2/2) + q_2 \sin^2 (\ta'_2/2), \non \\
0 &\simeq& p_2 \sin^2 (\ta'_2/2) + q_2 \cos^2 (\ta'_2/2). \label{z0condapp}
\end{eqnarray}
Moreover, in this regime we numerically find that $r_A \simeq 1$ and
$r \simeq 0$ for all $\theta_1$.

Next, we consider a change of $U_0=\mu_L +\de$ to $U_0=\mu_L-\de$ for
an arbitrary small value of $\de$ and a fixed applied bias
voltage $eV$. Using Eqs.~\eqref{ehwav1} and \eqref{reg2eqs}, it is easy
to see that under such a transformation $\theta_2 \Leftrightarrow
\theta'_2$ and $k_{z2} \Leftrightarrow k'_{z2}$. Thus, as long as $
r\simeq 0$, and $r_A \simeq 1$, Eqs.~\eqref{z0condapp} is
approximately invariant under this transformation with $p_1 \to
p'_1=p_2^{\ast}$, $q_1 \to q_1'=q_2^{\ast}$, $r \to r'=r^{\ast}$,
and $r_A \to r'_A = r_A^{\ast}$. Moreover, it is easy to see that the
same transformation keeps Eq.~\eqref{zdcond} invariant with $t_1 \to
t'_1 = t_1^{\ast} e^{-i \beta}$, $t_2 \to t'_2 = t_2^{\ast} e^{i
\beta}$ and $\nu \to \nu'=-\nu$. Thus the transmission probabilities
and hence the conductance $G_s$ (which is computed by integrating
over the azimuthal angle and is hence invariant under the change
$\nu \to -\nu$) remains approximately invariant under this
transformation. We note that, in contrast to the conductance of NBN
junctions, the invariance for $G_s$ is approximate and holds only for
$\theta_1, \theta'_1 \to 0$ for which $r_A \simeq 0$ and $r \simeq 1$.

Finally, we study the dependence of the subgap tunneling conductance on the
barrier strength $\chi$ in the thin barrier limit. In this limit $\ta_2,
\ta'_2, k_{3z}d, k'_{3z d} \to 0$ and $k_{2z}d, k'_{2z}d \to \chi= U_0 d/
\hbar v_F$. Substituting this in Eqs.~\eqref{z0cond} and \eqref{zdcond}, we
obtain
\begin{widetext}
\begin{eqnarray} \cos^2(\ta_1/2) + r \sin^2(\ta_1/2) &=& p_1 = [t_1 e^{i \beta}
\cos^2(\ta_3/2) + t_2 \sin^2(\ta_3/2)]e^{i \nu}, \non \\
\sin^2(\ta_1/2) + r \cos^2(\ta_1/2) &=& q_1 = [t_1 e^{i \beta} \sin^2(\ta_3/2)
+ t_2 \cos^2(\ta'_3/2)]e^{-i \nu}, \non \\
r_A \cos^2 (\ta'_1/2) &=& p_2 = [t_1 \cos^2(\ta_3/2) + t_2 e^{i \beta}
\sin^2(\ta'_3/2)]e^{i \nu}, \non \\
r_A \sin^2 (\ta'_1/2) &=& q_2 = [t_1 \sin^2(\ta_3/2) + t_2 e^{i \beta}
\cos^2(\ta'_3/2)]e^{-i \nu}, \label{tbcond} \end{eqnarray}
\end{widetext}
where $\nu= (n_1-n_2)\phi_{\vec k} -\chi$. These equations can be
solved to obtain the expression for $r$ and $r_A$. We note here that
$\chi$ enters these equations only as a constant shift to the
azimuthal angle $\phi_{\vec k}$. This ensures that $G_s$, similar to
its counterparts in spin-half Weyl and multi-Weyl semimetals, will
be an oscillatory function of the barrier strength $\chi$ if $n_1=n_2$;
in contrast, for $n_1\ne n_2$, $G_s$ becomes independent of $\chi$ in
the thin barrier limit. This behavior is shown in Fig.~\ref{fig8}.

\section{Discussion}
\label{diss}

In this work, we have studied the transport properties of pseudospin-one
fermions in the presence of a potential barrier. Such fermion
systems host quasiparticles which obey an effective spin-one Dirac
equation. Thus transport in NBN and NBS junctions show
unconventional features which are absent in similar junctions of both
conventional (Schr\"odinger) metals and pseudospin/spin-half Weyl semimetals.

One of the key features of ballistic transport in junctions hosting
pseudospin-one fermion is that for these junctions, current
conservation does not require continuity of all components of the
wave function across the junction. This feature can be contrasted
with Schr\"odinger materials where conservation current enforces
continuity of both the entire wave function and its derivative and
spin-half Dirac/Weyl semimetals where it enforces continuity of the
entire wave function. We show that this is a natural consequence of the
non-invertibility of the spin matrices $S_{\al}$ which generate the
spin/pseudospin algebra. This property is therefore expected to hold
for all integer spin/pseudospin Weyl systems where the expression
for the current $J_{\al}$ does not involve all the components of the
fermion wave function; indeed, for an integer spin $s$
Weyl fermion, one requires continuity of only $2s$ components of the
wave function. This features allows for current conservation without
imposing constraints on all components of the fermion wave function.

The most notable consequence of the discontinuity in some components
of the wave function is the collimation properties of transport
through such junctions. It is well-known that
pseudospin-one electromagnetic waves with effective Dirac-like
dispersion may show such a collimation in the presence of an array of
potential barriers \cite{fang1}; however, here we demonstrate
perfect collimation for a single barrier. We show that for both NBN
and NBS junctions of these materials, the transport is collimated
for a specific value of the barrier potential. It can be
analytically shown that any fermion that approaches the barrier of a
NBN junction with energy $\mu_L + eV$ at a finite angle of incidence
gets reflected off the barrier with unit probability if $U_0= \mu_L +
eV$. In contrast, a fermion approaching the barrier at normal
incidence is transmitted with unit probability. Since the latter
effect is a manifestation of Klein tunneling, this makes these
systems interesting platforms for observing Klein tunneling through
transport experiments. Similar effect occur for hole
mediated transport for $U_0=\mu_L -eV$. We also note that the NBN
junctions hosting pseudospin-one fermion system exhibit an
interesting symmetry of the conductance, namely, $G$ is the same for
two values of $U_0$ which are related to each other by reflection
about the value $\mu_L + (-) eV$ provided that the transport occurs
via motion of electron (hole)-like quasiparticles.

For NBS junctions, since the transport involves both electrons and
holes, such dips in the conductance signifying collimation is seen
for both $U_0=\mu_L +eV$ and $U_0=\mu_L -eV$. Such collimation does not
occur in spin-half Dirac/Weyl systems since, as shown in
Sec.~\ref{nbn}, it requires a discontinuity of the fermion wave
function across the junction and can thus occur only for junctions
hosting integer pseudospin Dirac fermions for which current
conservation does not enforce continuity of the entire wave
function. We also note that for these junctions,
where the transport is mediated by both electron- and hole-like
quasiparticles, $G_s$ is, in general, not invariant under $U_0 \to
2(\mu_L \pm eV)-U_0$ for any finite $V$. This is a consequence of
the participation of both electron- and hole-like quasiparticles in
transport. However, we find that in the regime where $eV \ll \mu_L$
and $U_0$, only channels corresponding to near-normal incidence of the
electrons contribute to $G_s$. In this regime, $r \simeq 0$ and $r_A
\simeq 1$ and the subgap tunneling conductance $G_s$ can be shown to
have an approximate invariance under the transformation of
$U_0=\mu_L + \de$ to $U'_0=\mu_L -\de$ for any fixed applied
voltage. Thus a plot of $G_s$ as a function of $U_0$ appears to be
almost reflection symmetric about $\mu_L$ as shown in Fig.~\ref{fig7}.

In contrast, the barrier potential dependence of spin-one Weyl
fermions is qualitatively similar to its spin-half Weyl
counterpart in the thin barrier limit ~\cite{sinha1}. We find that
in this limit, the tunneling conductance of NBN and NBS junctions of
these materials oscillates with $\chi$ for $n_1=n_2$; in contrast,
they become independent of $\chi$ if $n_1 \ne n_2$. The latter
phenomenon, also seen for a junction between spin-half Weyl and
multi-Weyl semimetals, constitutes a signature of the change in the
topological winding number of the system across the junction~\cite{sinha1}.

We note that our theoretical predictions can be easily tested in
experiments. Several materials are expected to be candidates for
pseudospin-one fermions~\cite{crystalrefs}. We predict that a NBN
junction of these materials will show dips in tunneling conductance
when the barrier potential is tuned to $\mu_L + eV$ (for electron
transport) or to $\mu_L -eV$ (for hole transport).
Moreover, we also expect $G/G_0$ to be identical for barrier
potential values $U_0$ and $2(\mu_L +(-) eV)-U_0$, where $+(-)$ sign
is applicable for electron (hole) mediated transport. To realize
this behavior experimentally, one needs, for electron transport, to
apply a potential $U_0$ which is close to $\mu_L$; thus these
experiments would be easier to perform in systems where the Fermi
energy of the pseudospin-one fermions is close to the Weyl nodes. In
this context, we also note that our theoretical analysis has been
carried out in the ballistic regime and assuming that there is no
internode scattering between the Weyl fermions. The former can be
justified by noting that in these systems (as shown for spin-half
Weyl and two-dimensional Dirac systems in
Ref.~\onlinecite{disorder1}), there is usually always a
quasi-ballistic regime at weak disorder where the analysis of the
ballistic junctions holds. The latter approximation can be justified
by noting that internode scattering is usually suppressed at low
energies~\cite{internode1}; moreover, they can only occur if the two
Weyl nodes occur at the same transverse momentum since scattering
from the barrier must conserve momentum.

In conclusion, we have studied ballistic transport in NBN and NBS junctions
of pseudospin-one Weyl fermions. We have shown that current conservation in
such junctions does not require continuity of the entire fermion wave
function. We have identified this property to be the reason for perfect
collimation in such junctions at specific values of the barrier potential.
We have discussed experiments which can test our theory.

\vspace{.8cm}
\centerline{\bf Acknowledgments}
\vspace{.5cm}

S.N. thanks D. Sinha for useful discussions. K.S. thanks J.D. Sau for
discussions. D.S. thanks DST, India for Project No. SR/S2/JCB-44/2010 for
financial support.

\end{document}